# Wakefields studies for the SXFEL user facility

Minghao Song[1,2], Chao Feng[1], Dazhang Huang[1], Haixiao Deng[1,*], Bo Liu[1], Dong Wang[1]

*1 Shanghai Institute of Applied Physics, Chinese Academy of Sciences, Shanghai, 201800, P. R. China*

*2 University of Chinese Academy of Sciences, Beijing 100049, P. R. China*

**Abstract:**

Besides the original seeded undulator line, in the Soft X-ray free-electron laser (SXFEL) user facility at Shanghai, a second undulator line based on self-amplified spontaneous emission is proposed to achieve 2 nm laser pulse with extremely high brightness. In this paper, the beam energy deviation induced by the undulator wakefields is numerically obtained, and it is verified to have a good agreement between 3D and 2D simulation results. The beam energy loss along the undulator degrades the expected FEL output performance. Impact of wakefields on pulse energy, radiation power and spectrum is discussed, as well as the benefits of compensation obtained with a taper in the undulator field. And using the planned SXFEL diagnostic, a longitudinal wakefields measurement experiment is proposed and simulated.

1. Introduction

The invention of free-electron laser (FEL) provides the cutting-edge scientific technique in various fields, ranging from unveiling chemical dynamics to probing surface catalysis, and from material science to medical physics [1, 2]. Nowadays, more and more X-ray FEL facilities are under construction and operation all over the world, such as LCLS [3, 4] and FLASH [5], which can generate extremely ultra-short pulse and high intensity radiation. Currently, the first X-ray FEL in China is under construction in Shanghai, namely SXFEL test facility [6]. To extend the SXFEL wavelength down to water window region of 2 nm, SXFEL user facility is proposed.

The basic layout of SXFEL user facility is described in Fig. 1. On the basis of test facility, the four more C-band accelerators will be installed in reserved LINAC space, which could further accelerate the electron beam energy from 0.84 GeV to 1.5 GeV. Thus, with the additional 7 undulator segments, the baseline seeded FEL scheme of test facility, i.e., two-stage of high gain harmonic generation (HGHG) [7] and echo-enabled harmonic generation (EEHG) [8] will achieve saturation at 3 nm wavelength. Meanwhile, to fully cover the water-window wavelength, an extracted FEL branch operated in self-amplified spontaneous emission (SASE) [9] mode will be built, which simply consists of the undulator segments and insert transitions.

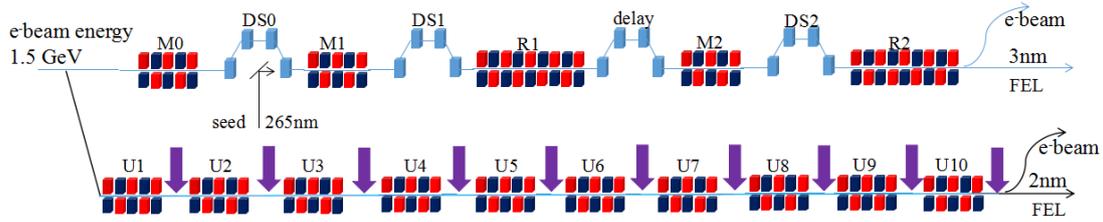

Fig. 1. The layout of SXFEL user facility consists of seeded and SASE branches.

SXFEL aims to produce high brightness photon pulse with narrow bandwidth. However, the designed FEL pulse may be degraded by negative wakefield effects along the SXFEL undulator section. Thus, calculations of wakefields in the undulator chambers and other impedance items are necessary to evaluate the FEL performance, e.g., in terms of pulse energy and FEL bandwidth. An intensive study of longitudinal wakefields and its effects on seeded FEL branch outputs for SXFEL test facility has been carried out [10], under Gaussian beam profile assumption. More recently, analytical descriptions and 2D computations of wakefields have been preliminarily considered for SXFEL user facility [11].

In general, the wakefields calculation relies on the analytical formulas and numerical codes. For the SASE branch of SXFEL user facility, besides the undulator vacuum chamber, however, the beam pipe will need to be interrupted 10 times by pipes with round cross section that connect each undulator. Therefore, it is necessary to check the 3D effects of the conventional solutions for wake calculations in those irregular structures, e.g., flat vacuum chamber and rectangular-to-round step-out. Thus, numerical calculations could be done with CST particle studio [12], a 3D computer solver program. In this paper, the wakefields generated in the resistive wall, surface roughness, and the discontinuities of beam-pipe are considered for SXFEL user facility, under the real beam profile tracked with ELEGANT [13], and a comparison of 3-D and 2-D wakefields calculation is presented. Considering the FEL lasing under the effects of wakefields, it is shown that, the pulse energy of the final 2 nm FEL pulse will be degraded by a factor of 7.5 with a gradual beam energy loss of 4 MeV along the whole undulator section. Fortunately, it can be compensated by a fine tuning of the undulator taper. In addition, a possible experiment of wakefields measurement at SXFEL user facility is proposed and discussed.

2. Wakefields calculation

In SXFEL user facility, linear accelerating section is made up of accelerating modules and relative simple structures. Thus consideration of wakefields calculation here is limited within undulator section.

For the SASE beamline of SXFEL user facility, 10 in-vacuum undulators will be used to reach the target wavelength. Then the electron beam should pass through the space between two opposite magnetic poles, which can be designated as flat vacuum chamber. Therefore, the short-range resistive wall wake effect arisen from metallic wall can be calculated by the formulas [14, 15]. The surface characteristics of vacuum chamber shows to be more complex under the observation of atomic microscopy [16] and surface roughness wake [17, 18] appears relatively small, thus the real shape could be simplified in terms of conductive wall with smooth surface only. Due to the limitation of CST wakefield solver, a Gaussian bunch current is considered.

In Fig. 2 (a), the comparison of resistive wall potentials for one segment of the undulator vacuum chamber is presented. It is found that the two results are correspondent well with each other despite of a slight deviation. And it rightly proves the accuracy of model built in CST and analytical formulas. For a Gaussian distribution with FWHM pulse duration of 250 fs, here, the mean beam energy losses due to resistive wall are 59.5 keV and 44.2 keV from the theory and CST simulation, respectively. Meanwhile, it has demonstrated that CST results are highly agreed on well with analytical consequences, in order to evaluate the real FEL performance at the end of undulator section, thus a wakefield calculation with real beam profile is needed to obtain the real energy loss. In light of the Fig. 2 (b), one can compute that the mean energy loss from the resistive wall in single undulator segment is approximately 106 keV.

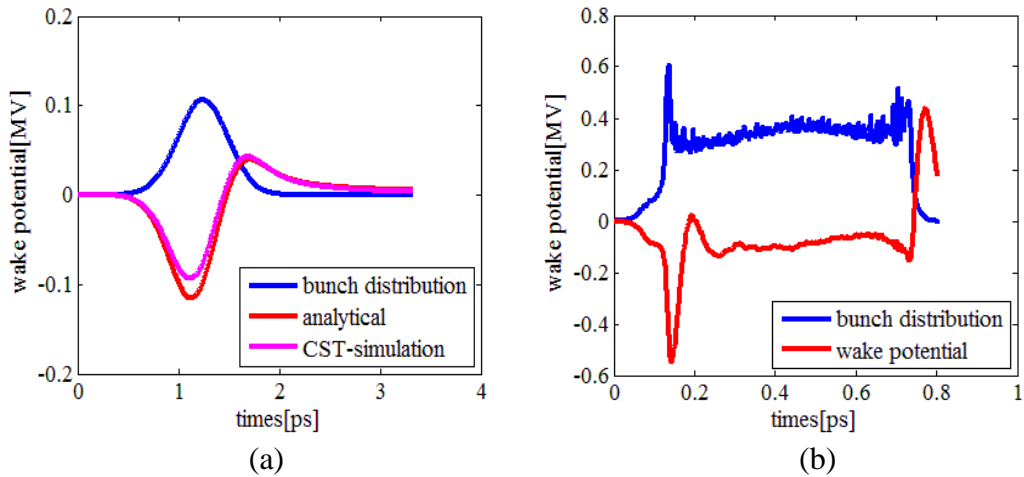

(a)        (b)

Fig. 2. (a) The comparison of the resistive wall between the theoretical analysis and CST simulation and (b) The wake potential under the real bunch profile.

It seems that SASE baseline shows slightly simpler compared with configuration of cascade HGHG baseline [11]. However, the required beam position monitors, profile monitors, quadruples and correctors are still needed between undulator segments to adjust the electron beam trajectory. Thus, the radiators need to be interrupted and connected by these transitions.

2-D Geometric wakefield has been studied carefully in [10] with the help of ABCI [19]. While due to the limitation of ABCI, a more accurate wakefield calculation of interaction between discontinuities and electron beam will be revealed here.

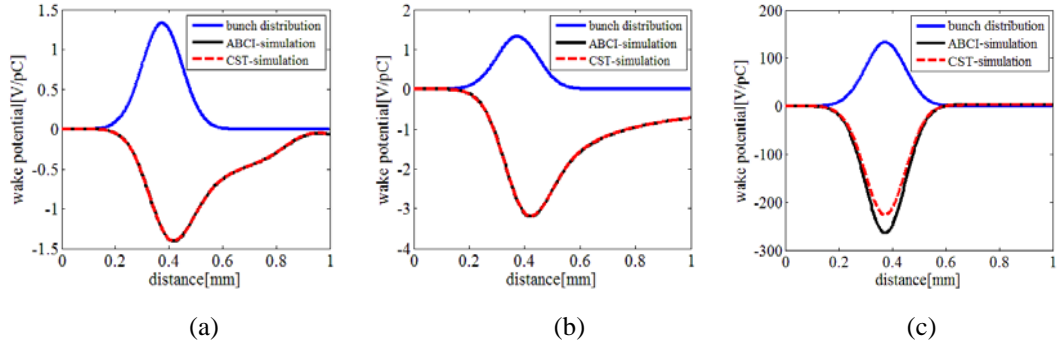

(a)    (b)    (c)

Fig. 3. The comparison of the wake potentials between ABCI and CST simulation. (a) Flange wakefield comparison, (b) Profile monitor wakefield comparison and (c) Step-out wakefield comparison.

From the CST and ABCI simulation, it can be found that the results of flange, profile monitor and step-out are almost the same with each other under Gaussian distribution. The mean beam energy losses are 0.12 keV, 1.12 keV and 77.06 keV on the basis of Fig. 3, respectively. It is found that the mean energy deviation of step-out is one order of magnitude, while the mean energy losses of flange and profile monitor are relatively small. Generally, the wakefield of undulator insertion could be obtained more accurately through simulating structure as a whole under the CST interface. However, the calculated meshes will be dramatically increased to several billions when applying for an ultra-short bunch like SXFEL user facility case. So due to the limitation of computation ability, as it is also demonstrated that ABCI has an excellent agreement with CST consequences, in this sense, which means a real 3-D structure wakefield computation can be approximately replaced by ABCI simulation. Thus the whole undulator transition would be taken into ABCI program under this circumstance, and wakefield shown in Fig. 4 can be generated considering adopting real bunch profile. According to the plot, it is not difficult to find that mean energy loss is 264.8 keV.

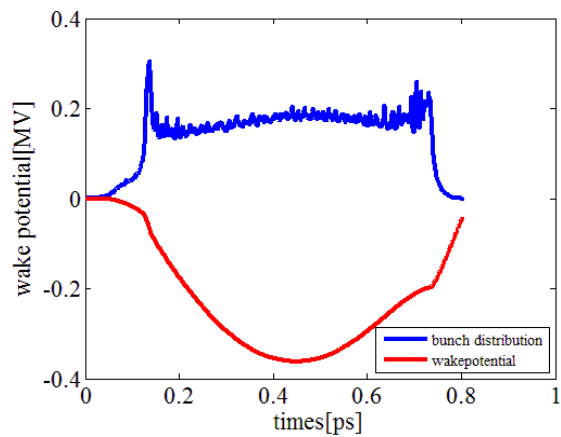

Fig .4. The wake potential of undulator transition with the real bunch profile.

As aforementioned, single undulator wakefield and transition wakefiled have been calculated under the real bunch profile. Here it is worth noting that one undulator wakefield calculation also contain surface roughness part. Furthermore, the total energy deviation can be obtained through a simple superposition of 10 undualtors and transitions, which is approximately 4 MeV, as shown in Fig. 5.

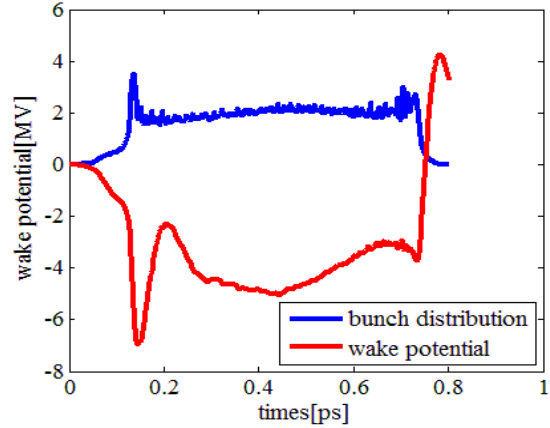

Fig. 5. The total wake potential of SASE undulator section.

### 3. Wakefield impact on photon pulse

Considering the wakefield effects in the undulator system, once the sliced beam energy loss becomes comparable to the pierce parameter, most electrons will not satisfy the resonant condition again, leading to FEL saturation and low FEL efficiency. Generally, the FEL reduction caused by the continuous beam energy loss can be compensated by gradually tapering the undulator to sustain the interaction even when the electrons lose a large fraction of their energy. Undulator tapering technique has been investigated since early days of FEL technology when the FEL was proposed to produce very high power [20, 21]. In this paper, only the linear taper of SXFEL undulator is discussed, defined as

$$K(z) = K_0 (1 - g)\frac{z}{L}$$

Where $K_0$ is the initial undulator parameter, $L$ is the total length of undulator, and g is the linear taper coefficient.

To model the expected SASE FEL performance of SXFEL user facility, start-to-end time-dependent simulation is used to evaluate the wakefield effects and the compensation strategy using taper undulator technique. The simulated 2 nm pulse energy at the exit of the undulator is 447 μJ without the wake effects. Fig. 6 displays the FEL pulse energy dependence on the taper coefficient when the wakefields are applied in the simulation. Considering that there exist different options for satisfying the required magnetic field of the in-vacuum undulator, two undulator gap cases (3 mm and 4 mm) are considered here. According to the starting points of this plot, it is interesting to find that a larger undulator gap contributes to a lower wakefields effects. However in the following discussion, the 3 mm gap case is studied. On the basis of these curves, one can find that tapered pulse energy at the exit of undulator section can be several times larger compared to the un-tapered cases, and especially the FEL pulse energy can be recovered to the no-wake situation, with a taper coefficient g=0.4%. In principle, if it is continued to increase the tapering coefficient, the larger pulse energy will be obtained. While such relevant research is beyond the scope of this paper, thus more detailed staff is not described anymore here.

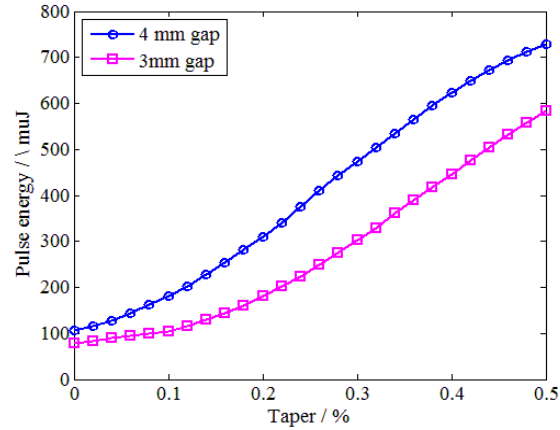

Fig. 6. The variation of pulse energy with the change of taper coefficient.

The real simulation results can be obtained through tracking electron beam from the start to the end of the undulator section. In order to avoid the shot-to-shot fluctuation of SASE FEL, 50 shots simulations with different shot noise are carried out for the interested cases here. Fig. 7 displays the predicted pulse energy evolution, FEL spectrum and instantaneous radiation power of 500-pC charge at around the 50 m in the undulator section.

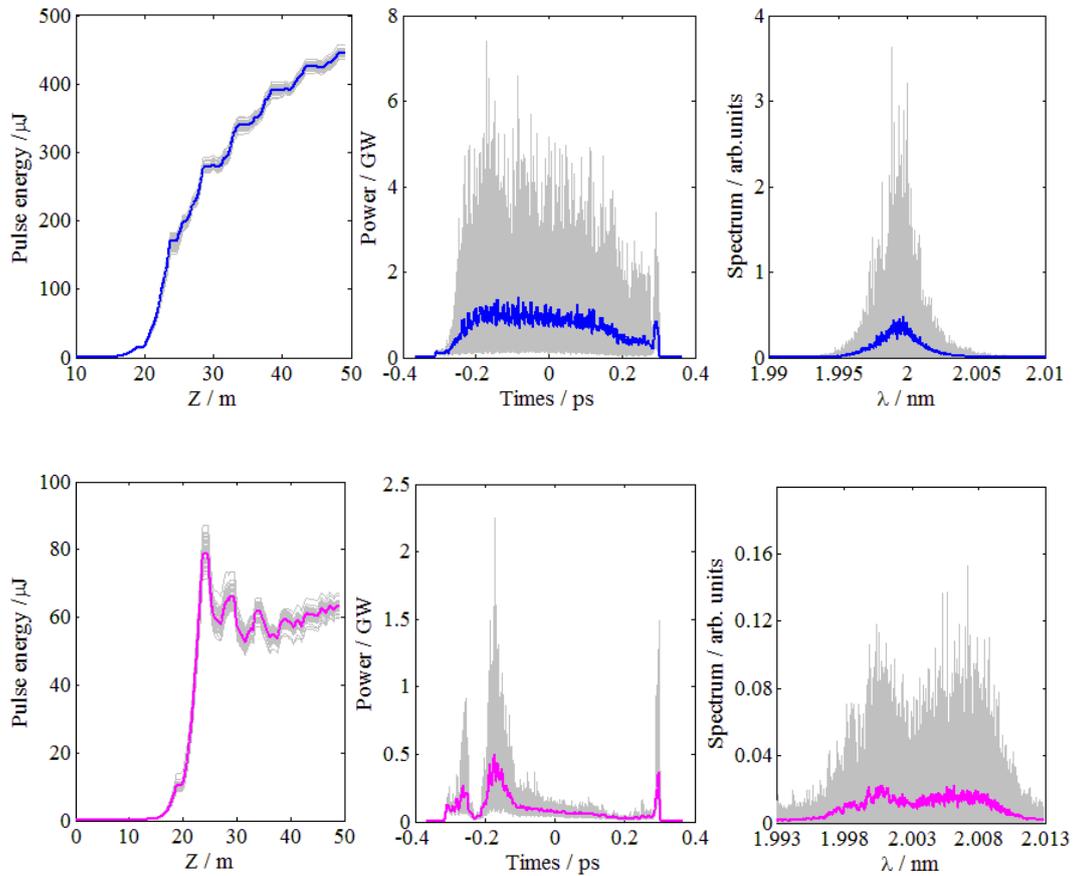

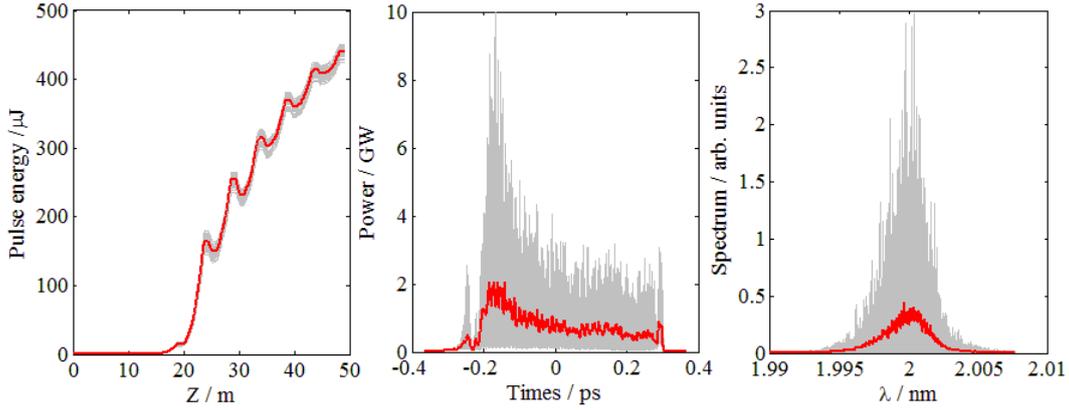

Fig. 7. The upper shows the evolution of output pulse energy along the undulator section, radiation power versus pulse duration and characteristic of spectrum under the condition of without wake effects. The middle part presents the case under the effects of wakefield. The lower plots illustrate the case for the undulator taper compensation despite of the wakefields.

The situation of without taking wakefields into account is presented in the upper results of Fig. 7, the steady increase of pulse energy before saturation is mainly dominated by the exponential gain and coherence building. One can find that the final photon pulse can achieve a maximum energy of nearly 450 μJ, moreover, the radiation power of the 2 nm FEL has an average level of approximately 1 GW in the absence of wakefield energy losses for a pulse duration of 0.8 ps. It is found that the comparison of radiation power profile with real bunch distribution shows that there is actually little FEL radiation generated in the head and tail high current spike regions. However, provided that the whole energy loss induced by the uncompensated wakefield is considered and illustrated in middle part of Fig. 7, it is obvious to find that pulse energy, radiation power along with spectrum all appear differently, in which the time-integrated pulse energy drops more than seven-fold compared with no-wake case and the overall radiation power level is strongly suppressed to a quarter of former case, consequently, the spectrum is also degraded and broadened under the effects of wakefields. In order to compensate the energy loss caused by wakefields to obtain expected FEL lasing, one should slightly tune the magnetic field of each undulator to match the FEL resonance. It is obviously found that approximately 0.4% taper coefficient is required to compensate output average power on the basis of Fig. 6 for the situation of undulator gap equal to 3 mm. The lower plots describe the case of wake introduced with taper compensation. Eventually, it is deserved to note that pulse energy, radiation power and spectrum all can be approximately retrieved to the no-wake case through a process of optimization.

4. **Proposed wakefield measurement**

Currently, wakefields are frequently considered in the beam control and accelerator operation. The difficulty of wakefield measurement has been existing in circular machine for many years. It is now equally important for the FEL facility due to stringent requirement of extremely high brightness. Radio frequency deflectors are widely used for time-resolved electron beam energy, emittance and radiation profile measurements in modern FEL facilities [22, 23]. With SXFEL user facility diagnostic beamline which has been studied in [24], it has demonstrated that bunch longitudinal profile can be reconstructed from image of the observation screen. Therefore, the whole longitudinal wakefield arisen from undulator section could also be reconstructed with the help of RF deflecting cavity installed in beamline.

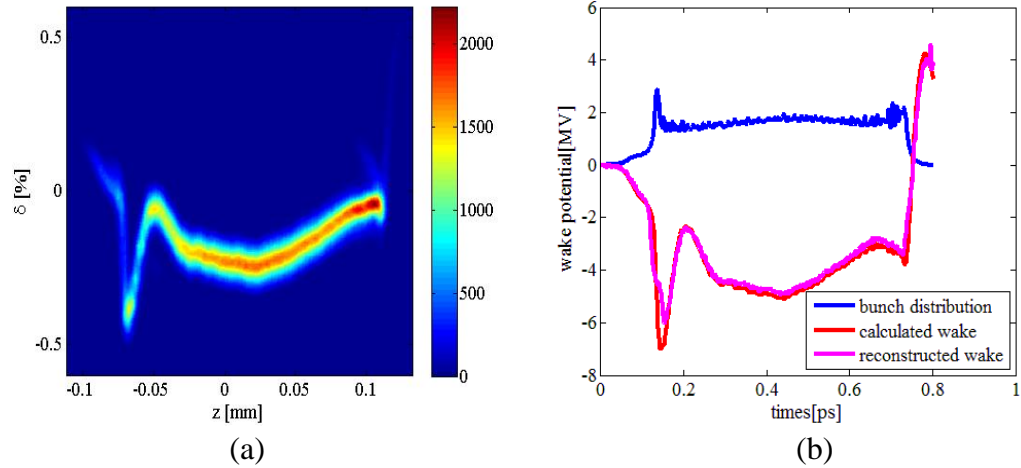

Fig. 8. Measurement consequences under the case of undulator gap equal to 3 mm. (a) The longitudinal phase space on the observation screen and (b) Comparison between reconstructed wake and calculated wake.

Figure 8 illustrates the longitudinal phase space on the observation screen and comparison between the reconstructed wakefields and the calculated wakefields. Generally, the longitudinal phase space on the observation screen at the exit of the undulator has taken wakefields into account in the real situations, which could be expressed by the Fig. 8 (a). To reconstruct the wakefields in the undulator, it could be achieved by comparison between the phase space on the screen before and after the undulator. In accordance with the Fig. 8 (b), it can be found that there is a good agreement between these two consequences, in this sense, it can also demonstrate that the time-resolved resolution of optimized SXFEL diagnostic beamline satisfy the experiment's requirements.

In this paper, the undulator gap of 3 mm is only calculated and simulated, to better understand the influence of gap variation on final wakefields, one needs to consider the wake effects under a wide range of undulator gaps. Here the total energy losses with gap from 3 mm to 12 mm are computed. Obviously, the induced energy deviations will experience a descending trend with the increasing gap, for instance, the energy loss is approximately 1 MeV when the gap equals to 12 mm, which is a quarter of 3 mm case and hardly have impact on the FEL performance compared with electron bunch central energy of 1.5 GeV. At the same time, to examine the accuracy of measured results, the comparison between computed energy loss and measured consequence is shown in Fig. 9, as previously mentioned, the two results are close to each other due to a good agreement in wake potential distribution.

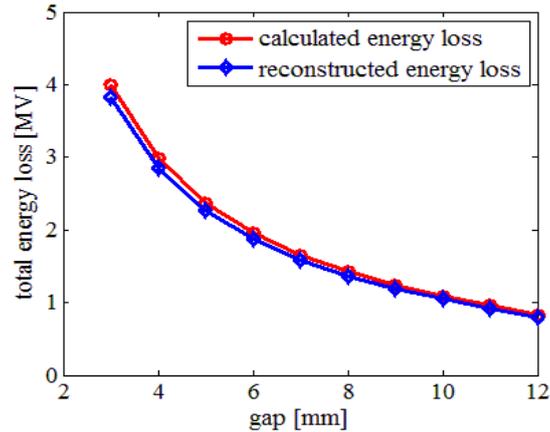

Fig. 9. Consequence of calculated energy loss and reconstructed energy loss with different undulator gaps as well as comparison between them.

In further study, apart from the longitudinal wakefields measurement and investigation, transverse beam kick measurements are also a real concern when preserving the low emittance and orbit correction in modern FEL machines [25]. In combination with a sub-micron resolution cavity beam position monitor [26] that consist of reference cavity and position cavity, the position measurement system of SXFEL user facility diagnostic beamline provides the ability required for measuring wake-induced beam offsets and allows us to characterize the transverse wakefield source.

5. Conclusion

Generally, on account of the relative long electron bunch and simple accelerating structure, wakefield effects in the linear accelerator is small, and more importantly could be compensated with the help of feedback system. Therefore, in this paper, besides the theoretical analysis and 2-D ABCI simulation, a more accurate 3-D wakefields calculation using the CST particle studio environment is accomplished for the resistive wall, surface roughness and geometric discontinuities. Moreover, it is demonstrated that there is a reasonable agreement between 3-D results and 2-D consequences. Thus, in the start-to-end FEL simulation with the wakefields effects, the total beam energy loss of 4 MeV along the whole undulator section causes the serious degradation of FEL performance, for example, the pulse energy and average radiation power will fall seven times. Fortunately, the FEL performance loss caused by wakefields could be compensated by a linear taper in the undulator field. With the optimum taper, the losing FEL could be recovered which is verified in the simulation. Furthermore, it is interesting to find that the output energy will experience an approximate linear fit under the circumstance of increasing taper coefficient, which can help decide an appropriate coefficient for target pulse energy. What's more, in the SXFEL user facility, wakefields measurement and calibration could be done under the optimized TDS diagnostic beamline. In the simulation, the final reconstructed results have a good agreement with the given ones. More simulations and experiments will be accomplished in the future to help understanding the longitudinal and transverse wakefields in modern FEL facilities.

**Acknowledgement**

The authors are grateful to Xiao Hu, Wei Zhang, Meng Zhang, Kai Li and Han Zhang for helpful discussions and useful comments. This work was partially supported by the National Natural Science Foundation of China (11475250 and 11322550) and Ten Thousand Talent Program.